\documentclass[aps,prd,twocolumn,superscriptaddress,nofootinbib,showpacs,preprintnumbers]{revtex4}

\usepackage{amsmath,amssymb,amsfonts}
\usepackage{graphicx}
\usepackage{graphpap} % for numbered coordinate grid

\usepackage{hyperref} % for HyperTeX cross-referencing
\usepackage{subfigure}
\def\be{\begin{equation}}
\def\ee{\end{equation}}
\def\ba{\begin{eqnarray}}
\def\ea{\end{eqnarray}}
\def\nn{\nonumber}

\def\a{\alpha}

\def\g{\gamma}
\def\d{\delta}

\def\k{\kappa}

\def\m{\mu}

\def\p{\pi}

\def\r{\rho}
\def\vr{\varrho}

\def\D{\Delta}

\def\L{\Lambda}

\def\Vz{V_0}

\newcommand{\abs}[1]{{\left|{#1}\right|}} % abs (variant delimiters)
\newcommand{\tp}{\widetilde{p}} %tilde p
\newcommand{\tc}{\widetilde{c}}   % tilde c

\newcommand{\Pl}{\ell_{\rm Pl}} % Planck length
\newcommand{\mubar}{{\bar \m}} % mu bar
\newcommand{\K}{{\cal K}}
\newcommand{\Ham}{{\mathcal H}}

\newcommand{\secref}[1]{Section~\ref{#1}}

\newcommand{\eqnref}[1]{(\ref{#1})}

\newcommand{\appref}[1]{Appendix~\ref{#1}}

\begin{document}

\preprint{IGPG-07/5-1}

\title{The behavior of non-linear anisotropies in\\
bouncing Bianchi I models of loop quantum cosmology}
\author{Dah-Wei Chiou}
\email{chiou@gravity.psu.edu}
%\affiliation{Department of Physics,
%University of California at Berkeley,\\
%Berkeley, CA 94720,
%U.S.A.}
\affiliation{Institute for Gravitational Physics and Geometry,\\
Physics Department, The Pennsylvania State University,\\ University Park, PA 16802, U.S.A.}
\author{Kevin Vandersloot}
\email{Kevin.Vandersloot@port.ac.uk}
\affiliation{Institute for Gravitational Physics and Geometry,\\
Physics Department, The Pennsylvania State University,\\ University Park, PA 16802, U.S.A.}
\affiliation{Institute for Cosmology and Gravitation,\\ University of Portsmouth,\\
Portsmouth, PO1 2EG, UK}

%\date\today

\begin{abstract}
In homogeneous and isotropic loop quantum cosmology, gravity can behave repulsively at Planckian energy densities leading to the replacement of the big bang singularity with a big bounce. Yet in any bouncing scenario it is important to include non-linear effects from anisotropies which typically grow during the collapsing phase. We investigate the dynamics of a Bianchi I anisotropic model within the framework of loop quantum cosmology. Using effective semi-classical equations of motion to study the dynamics, we show that the big bounce is still predicted with only  differences in detail arising from the inclusion of anisotropies. We show that the anisotropic shear term grows during the collapsing phase, but remains finite through the bounce. Immediately following the bounce, the anisotropies decay and with the inclusion of matter with equation of state $w < +1$, the universe isotropizes in the expanding phase.
\end{abstract}


\maketitle

\section{Introduction}
Recently, the investigation of loop quantum cosmology (For a review see \cite{Bojowald:2006da}) in homogeneous
and isotropic universes has indicated that the classical big-bang singularity
can be replaced with a big-bounce \cite{Ashtekar:2006rx,Ashtekar:2006uz,Ashtekar:BBII,Newkplus1,Vandersloot:2006ws}. In these scenarios
gravity can be interpreted as becoming repulsive in the  Planckian
high energy regime, implying that our
current expanding universe would have been preceded by a contracting phase.
The presence of a big-bounce has been shown to be a rather generic
and genuine feature of the quantum gravitational effects of loop quantum cosmology
and does not require any exotic matter which violates  energy conditions \cite{Singh:2006im}.
An exciting fact of bouncing cosmological models is that the scales measured in
the cosmic microwave background (CMB) can be in causal contact if
the current expanding phase is preceded by a contracting one, thus opening the possibility
for a replacement of the standard inflationary scenario.

In order to more fully develop this scenario, one must go beyond the assumption
of homogeneity and isotropy. It is both the inhomogeneities and anisotropies
that are expected to grow in a collapsing phase and thus a proper
accounting of these fluctuations is required. One particular question that
immediately arises is whether the presence of the bounce is
stable under the inclusion of inhomogeneities and anisotropies.
A proper description of the inhomogeneous perturbations and anisotropies
would then provide an answer to the question of whether a suitable alternative
to inflation can be constructed, and/or what possible cosmological signatures
may result.

In this paper we do not seek to answer all these questions as work
on including inhomogeneities in loop quantum cosmology is in its infancy (initial
progress can be found in \cite{Bojowald:2006tm,Bojowald:2006qu,Bojowald:2006zi,Bojowald:2006zb}). Instead, we will focus on the
 behavior of anisotropies, studying the dynamics of the anisotropic Bianchi I model
in the framework of loop quantum cosmology.
In the classical Bianchi I universe sourced with matter with zero anisotropic stress,
the anisotropic shear term behaves as an effective matter component with energy density that scales as $a^{-6}$ in the Friedmann equation
with $a$ being the mean scale factor. Thus for matter with equation  of state $w<+1$,
the anisotropies will  dominate the collapsing phase as the singularity is approached. If the current
expanding phase of the universe was preceded by a collapsing phase, the inclusion
of anisotropies can be expected to play a significant result near the bounce.

The loop quantization
of the anisotropic Bianchi I model was initially studied in \cite{Bojowald:2003md} and
more recently in \cite{Chiou:2006qq}.  In this paper, we will study the dynamics
of the model at the level of
effective classical equations of motion that incorporate quantum effects
arising from the loop quantum Einstein equations. At the level of the
effective equations we study,
we show that
the big-bounce is indeed robust under the inclusion of non-linear anisotropies.
We will show that the anisotropic shear remains finite through the bounce and that
 if matter with equation of state $w<+1$ is included, the universe isotropizes
in the expanding phase. The results represent evidence that the bouncing scenario
of loop quantum cosmology is robust when the assumptions of homogeneity and isotropy
are relaxed and gives hope that the same can be said when inhomogeneous perturbations
are properly included in the theory.

\section{Classical Dynamics}\label{classical theory}
We start with the classical setup for the Bianchi I model. Loop quantum cosmology
is based on a Hamiltonian formulation and thus we will define the Hamiltonian
for the gravitational plus matter degrees of freedom. Hamilton's equations of
motion then are equivalent to Einstein's equations for the model considered.
Since the starting point of the quantization is the Hamiltonian, it is there that
we will incorporate the quantum effects in a modified effective Hamiltonian to be
introduced in the next section.

Starting with the dynamical variables that comprise the classical phase
space, we have three triad variables
 $\tp_I$ and three connection variables $\tc_I$ with
$I=1,2,3$. The classical metric given in terms of the
directional scale factors $a_I$ is
\ba
	ds^2 = -N^2\, dt^2 + a_1^2 \,dx^2 + a_2^2\, dy^2 + a_3^2 \,dz^2
\ea
with $N$ representing the lapse which is a freely specified function representing
the freedom to redefine the time variable, and
the coordinates $x,y,z$ are all valued on the entire real line (we are
not considering a compactified model). The triad variables
are directly related to the scale factors as
\be\label{eqn:p and a}
|\tp_1|=a_2a_3,\quad
|\tp_2|=a_1a_3,\quad |\tp_3|=a_1a_2.
\ee
Thus the triad variables encode information about the spatial geometry. The connection
$\tc_I$ will encode information about the curvature (essentially time derivatives of
the scale factors) as will be evident once Hamilton's equations are solved.

In the Hamiltonian framework, the classical equations of motion are derived
from a Hamiltonian obtained by inserting the homogeneous phase space
variables into the Hamiltonian of general relativity making it a functional
of $\tp_I$ and $\tc_I$ in this case. However, a non-triviality arises
from the spatial integrations in the Hamiltonian. Because of homogeneity,
these integrations diverge since we are considering the non-compact Bianchi
I model. To overcome this, the spatial integrations can be restricted to a finite
sized fiducial cell with
fiducial volume $\Vz = \int d^3 x$. We can use the fiducial cell to define untilded
variables $p_I, c_I$ as
\ba \label{eqn:puntild}
	p_I = \Vz^{2/3} \, \tp_I, \qquad c_I = \Vz^{1/3}\, \tc_I \,.
\ea
In the classical theory, one can freely rescale the coordinates $x,y,z$ while
leaving the physics invariant. Under this rescaling, one can show that the untilded
variables $p_I, c_I$ are invariant. Since the quantization is based on the untilded
variables, the quantum theory is manifestly invariant under this coordinate rescaling.
This is entirely analogous to the non-compact isotropic $k=0$ \cite{Ashtekar:2003hd} and $k=-1$ \cite{Vandersloot:2006ws} models where the same procedure is used.

With the understanding that the spatial integrations in the Hamiltonian are restricted
to the fiducial cell, the total Hamiltonian of the model is given by
\ba\label{eqn:cl Hamiltonian}
\Ham&=&\Ham_{\rm grav}+\Ham_{\rm matter}\nn\\
&=&\!\!\frac{- N}{\k\g^2\sqrt{{p_1p_2p_3}}}
\left(c_2p_2c_3p_3\!+\!c_1p_1c_3p_3\!+\!c_1p_1c_2p_2\right)\nn \\
&&+\Ham_{M}.
\ea
with $\Ham_M$ being the matter Hamiltonian.
Here we have $\k=8 \pi G$, and  $\g$ is known
as the Barbero-Immirzi parameter and represents a quantum
ambiguity of loop quantum gravity which is a non-negative
real valued parameter. Two sets of equations of motion derived
from the Hamiltonian
then govern the dynamics. First, the Hamiltonian of gravity
and matter is of the constrained type whereby it vanishes identically for
solutions to Einstein's equations. Thus an equation
of motion is given by
\be\label{eqn:Ham=0}
\Ham = 0.
\ee
Second, Hamilton's equations give the time evolution
of any phase space variable through the Poisson bracket:
\be
	\dot{p_I} = \{ p_I, \Ham\},
\qquad
    \dot{c_I} = \{ c_I, \Ham\}.
\ee
The Poisson structure of the gravitational variables
leads to the only non-vanishing Poisson brackets
\be \{c_I,p_J\}=\k\g\,\d_{IJ}.
\ee
With this, Hamilton's equations for $c_I$ and $p_I$ are given by
\be\label{eqn:cl Hamilton's eq}
	\dot{p_I} = -\k \g \, \frac{\partial\Ham}{\partial c_I},
\qquad
    \dot{c_I} = \k \g \, \frac{\partial\Ham}{\partial p_I}.
\ee

To proceed further with the equations of motion, we must specify the
matter Hamiltonian. We take it to be of form
\be	\Ham_{M} = N \sqrt{p_1 p_2 p_3}\ \rho_M
\ee
with $\rho_{M}$ being the matter energy density. In
this paper, we assume that the matter has zero anisotropic
stress which implies that $\rho_M$ couples to $p_I$ in the
form
\be \label{rhomassum}\rho_M(p_1, p_2, p_3) = \rho_M(p_1 p_2 p_3) \,.
\ee
This assumption is true for scalar fields and perfect fluids which we will concentrate
on in this paper. Additionally, to derive the equations of motion we must specify
the form of the lapse. For simplicity of the resulting equations,
let us choose a form given by
\be
N=\sqrt{p_1 p_2 p_3}\,,
\ee
which means we are using a different time $t'$ variable than the usual cosmic
time $t$ given by
\be
dt'=(p_1 p_2 p_3)^{-1/2}dt.
\ee
Note that the choice of lapse is arbitrary and none of the physical results depend
on the choice.

We now derive the classical equations of motion. The Hamiltonian
with our choice of lapse is
\ba \label{Ham}
\Ham &=&-\frac{1}{\k \g^2}
\left({c_2c_3}{p_2p_3}+{c_1c_3}{p_1p_3}+{c_1c_2}{p_1p_2}\right)\nn\\
&&
+\,p_1p_2p_3\,\r_M.
\ea
The first set of Hamilton's equations in \eqnref{eqn:cl Hamilton's eq} for the time evolution of the triad then yields for instance
\be\label{eqn:cl dp/dt'}
\frac{dp_1}{dt'}=\frac{p_1}{\g}\left({c_2}{p_2}+{c_3}{p_3}\right).
\ee
Using the relations between the triad components and scale factors \eqnref{eqn:p and a}
and \eqnref{eqn:puntild}, we can solve these equations for the connection
coefficients $c_I$ to get
\be\label{eqn:cl a dot}
c_I=  \g \Vz^{-2/3}(a_1a_2a_3)^{-1}\frac{da_I}{dt'}\equiv\g\, \Vz^{1/3} \,\frac{da_I}{dt}.
\ee
These relations therefore provide us with the interpretation of the connection
components as containing information about the curvature, which in the the Bianchi I model is entirely
encoded as the extrinsic curvature here proportional to $da_I / dt$.

The next set of Hamilton's equations in \eqnref{eqn:cl Hamilton's eq} for the time evolution of the connection yields for instance
\ba\label{eqn:cl dc/dt'}
\frac{dc_1}{dt'}&=&-\frac{c_1}{\g}\left(c_2p_2+c_3p_3\right)\nn\\
&&+\k\g\, p_2p_3\left(\rho_M+p_1\frac{\partial \rho_M}{\partial p_1}\right).
\ea
Combining \eqnref{eqn:cl dp/dt'} and \eqnref{eqn:cl dc/dt'} gives
a key relation
\be\label{eqn:cl dpc/dt'}
\frac{d}{dt'}(p_Ic_I)=
\k\g\, p_1p_2p_3\left(\rho_M+p_I\frac{\partial\rho_M}{\partial p_I}\right).
\ee
It is here we use our assumption that the matter has zero anisotropic stress \eqnref{rhomassum}. This assumption implies $p_I\partial\rho_M/\partial p_I=p_J\partial\rho_M/\partial p_J$ and thus \eqnref{eqn:cl dpc/dt'} yields
\be\label{eqn:cl key relation}
\frac{d}{dt'}\left(p_Ic_I-p_Jc_J\right)=0,
\ee
which can be integrated to give
\be \label{aIJ}
	p_I c_I-p_J c_J = \g \Vz \, \a_{IJ}
\ee
with $\a_{IJ}$ being a constant anti-symmetric $3\times3$ matrix
satisfying by construction $\a_{12}+\a_{23}+\a_{31}=0$ and
the factors of $\g, \Vz$ are chosen for convenience.
Written in terms of the scale factors
this implies
\be \label{Hsol}
	H_I - H_J = \frac{\a_{IJ}}{a_1 a_2 a_3}
\ee
for the directional Hubble rates (in terms of cosmic time $dt=\Vz a_1 a_2 a_3\, dt'$
)
\be \label{HI}
	H_I \equiv \frac{\dot{a_I}}{a_I}.
\ee

Using \eqref{Hsol} and the vanishing of the Hamiltonian, we can
write down a generalized Friedmann equation. The vanishing of
the Hamiltonian \eqref{Ham} gives in terms of the Hubble rates
\be \label{rsol}
	H_1 H_2  + H_1 H_3 + H_2 H_3 = \k \r_M \,.
\ee
Let us next define
the mean scale factor $a$ as
\be
	a = (a_1 a_2 a_3)^{1/3},
\ee
from which the following relation holds
\be \label{adot}
	\frac{\dot{a}}{a} = \frac{1}{3} (H_1 + H_2 + H_3) \,.
\ee
This relation implies
\ba
	\bigg(\frac{\dot{a}}{a}\bigg)^2 &=& \frac{1}{3}(H_1 H_2  + H_1 H_3 + H_2 H_3) \nn \\
	+\frac{1}{18} && \!\!\!\!\!\!\!\!\!\!\!\!\big[ (H_1\!-\!H_2)^2 + (H_1\!-\!H_3)^2 +(H_2\!-\!H_3)^2 \big].
\ea
Finally, using \eqref{Hsol}
and \eqref{rsol}, the generalized Friedmann equation becomes
\be \label{Friedeqn}
	\bigg(\frac{\dot{a}}{a}\bigg)^2 = \frac{\k}{3} \r_M + \frac{\Sigma^2}{a^6},
\ee
where $\Sigma$ is given in terms of the constants of motion $\alpha_{IJ}$ as
\be
	\Sigma^2 \equiv   \frac{1}{18} \big( \alpha_{12}^2 + \alpha_{23}^2
	+ \alpha_{31}^2 \big) \,.
\ee
The \emph{anisotropic shear} $\sigma_{\mu \nu}\sigma^{\mu \nu}$ is related to $\Sigma$ through
\ba \label{shear}
	\sigma_{\mu \nu} \,\sigma^{\mu \nu} &\equiv&
	\frac{1}{3 } \bigg( (H_1\!-\!H_2)^2 + (H_2\!-\!H_3)^2 + (H_3\!-\!H_1)^2
	\bigg) \nn \\
	&=& \frac{6 \Sigma^2}{a^6}.
\ea
The equations of motion for $a(t)$ can now be determined from
the generalized Friedmann equation \eqref{Friedeqn}. The time evolution
for the directional scale factors $a_I(t)$ can then be determined from
a combination of \eqref{Hsol} and \eqref{adot}, once $a(t)$ is known.

Let us discuss the interpretation of the generalized Friedmann equation
\eqref{Friedeqn}. We first recognize the isotropic matter term $\k \r_M / 3$ on
the right hand side with the anisotropic shear term behaving as a stiff fluid with equation
of state $w=+1$. The isotropic limit is achieved when $\alpha_{IJ}=0, \Sigma = 0$ which
from the relations \eqref{Hsol} implies that the directional Hubble rates $H_I$
are identical indicating isotropic expansion.
Because the shear
terms scales as $a^{-6}$, for typical forms of matter
the early universe can be anisotropy dominated, with matter then dominating
the later stages. In particular, if $w < +1$, the later stages of expansion
tend to a more isotropic state, whereby the Hubble rates become identical
in all directions. Oppositely, as the singularity is approached, the universe
behaves like a vacuum universe as the shear term dominates the matter
term and Kasner like behavior occurs. This behavior is desribed in more detail
in appendix \ref{app1}.

As an example, let us consider a dust filled ($w=0$) Bianchi I universe with matter
density $\r_M = \tilde{A}/a^3$.  The generalized Friedmann equation can be solved analytically
giving
\be
	a(t) = \bigg[ \frac{3 \k \tilde{A}}{4}\, t^2 + 3 \Sigma t \bigg]^{1/3},
\ee
which gives the standard dust filled isotropic behavior at late times $a \propto t^{2/3}$
while at early times is anisotropy dominated $a \propto t^{1/3}$ and is
singular at $t=0$. The time evolution of the individual scale factors can
then be determined from \eqref{Hsol} and \eqref{adot}, which
gives for instance
\be \label{a1}
	a_1(t) = \bigg( \frac{3 \k \tilde{A}}{4}  + \frac{3 \Sigma }{ t} \bigg)^{\frac{-\alpha_{12}+\alpha_{31}}{9 \Sigma}} a(t).
\ee
Let us consider the situation where initially $a_1$ is contracting with the other two
directions expanding. In this case we have $\alpha_{12} <0$ and $\alpha_{31} > 0$.
It is easy to see from \eqref{a1}, that at late times $a_1(t)$ behaves as $a_1(t) \approx
a(t) \propto t^{2/3}$ and thus is {\em expanding} in accordance with the
isotropic behavior. At early times, however, it is
not too difficult to show that \eqref{a1} implies contraction ($\dot{a}_1 < 0$) and thus the scale factor
bounces at some finite time and eventually the universe isotropizes to
a matter dominated phase. Therefore with matter with equation of state $w < +1$, the universe
can initially be in a Kasner like epoch with one direction contracting and the other two expanding; eventually, however, such a universe will isotropize giving expansion in all
three directions.

Thus, at the classical level the anisotropies are expected to play an
important role as the singularity is approached.
In the isotropic setting of loop quantum cosmology, the bouncing
scenario has been well described in terms of an effective Friedmann
equation of the form \cite{Ashtekar:BBII}
\be
	H^2 = \frac{\kappa }{3} \rho_M ( 1- \frac{\rho_M}{\rho_{\rm crit}})
\ee
with $\rho_{\rm crit} \approx .82 \rho_{\rm Pl}$ being a critical energy
density of Planckian order. When the matter density reaches the critical
density, $H=0$ indicating a bounce. Because the bounce
occurs in the high energy regime near the classical singularity, it is
important to include the effects of the anisotropies. In the next
section we will answer the question as to what role the anisotropies
play in the bouncing scenario.

\section{Effective loop quantum dynamics} \label{effective theory}
As we have stated, the quantum modifications due to loop quantum
cosmology that we study exhibit themselves in the form of a modified effective
Hamiltonian. Since we study effective {\em classical} equations
of motion, we are by definition ignoring certain quantum degrees
of freedom. Properly speaking, in the quantum theory, dynamics
is understood through expectation values of  observables
calculated from semi-classical
wave packets. Thus additional effects which we do not study
can arise from features of the semi-classical state such as the spread
and so forth. The effective quantum modifications we study to first order
are insensitive to the features of the semi-classical wavefunction and are expected
to be valid provided the wavepacket remains sharply peaked though the evolution.
That this is a good approximation has been verified in the isotropic
models of LQC sourced with a massless scalar field\cite{Ashtekar:BBII,Newkplus1,Vandersloot:2006ws},
where the quantum dynamics have been extensively developed and
understood. Note that the effective equations we consider are somewhat heuristically
motivated and more systematic approaches to deriving effective equations
have been considered \cite{Bojowald:2006gr}, but ultimately proper justification
of any effective scheme requires the study of the quantum dynamics.

The main quantum effect arises from the fact that in
LQC, the connection variables $c_I$ do not have direct
quantum analogues and are  replaced by
holonomies (roughly exponentials of the connection).
This manifests itself in the effective Hamiltonian by replacing
the classical $c_I$ terms with sine functions
\be
	c_I \longrightarrow \frac{\sin(\mubar_I c_I)}{\mubar_I},
\ee
where $\mubar_I$ are real valued functions of the triad coefficients $p_I$
which are a  measure of the discreteness in the quantum theory. Classical
behavior is expected in the limit when $\mubar_I c_I \ll 1$ whence
$\sin(\mubar_I c_I) / \mubar_I \approx c_I$.

In the original construction of LQC, the $\mubar_I$ were taken to
be constants (referred to as $\mu_0$)\cite{Bojowald:2003md}; however it has been
shown that in the isotropic case this can lead to the wrong semi-classical
limit\cite{Ashtekar:BBII}. In the isotropic case it has been argued that
$\mubar$ should not be a constant, but should scale as $\mubar \propto 1/\sqrt{p}$
which has been shown to have a nice semi-classical limit\cite{Ashtekar:BBII}.
Extending this
scheme to the Bianchi I model is slightly more ambiguous as several
possibilities exist. We choose in this paper to focus on the scheme
proposed in \cite{Chiou:2006qq} where $\mubar_I$ are given by
\be \label{mubar}
\mubar_I=\sqrt{\frac{\D}{p_I}}\ ,
\ee
where $\D=\frac{\sqrt{3}}{2}(4\p\g\Pl^2)$ is the \emph{area gap} in the full theory
of LQC and $\Pl\equiv\sqrt{G\hbar}$ is the Planck length. An alternative scheme would be to have $\mubar_I \propto 1/a_I$ which
we will briefly discuss in \appref{altmubar}.
However the main physical results we will describe do not depend sensitively
on either of the scheme; only the quantitative results would change.
Better input from the full theory might provide justification for either scheme.
One advantage of the scheme \eqref{mubar} is that it is more amenable
to the study of dynamics with semi-classical states in the quantum theory, as
the resulting difference equation is much simpler. This would allow for
the test of the validity of the results presented here, by examining
semi-classical state behavior in the quantum theory.

Additional modifications due to loop quantum cosmology have been
studied extensively in the isotropic setting and pertain to operator
eigenvalues of inverse $p_I$ factors that would appear in the matter part of the
effective Hamiltonian. In the isotropic setting this amounts
to replacing factors of $p^{-3/2}$ in the matter Hamiltonian by an operator eigenvalue function
$d_j(p)$ which is bounded and vanishes at $p=0$ corresponding to
the classical singularity.\footnote{Similar inverse triad effects appear in the gravitational Hamiltonian leading to a function $s_j(p)$ \cite{Vandersloot:2005kh}. These corrections have been
typically been ignored, but see \cite{Magueijo:2007wf} for work on some cosmological implications.} Additionally an ambiguity parameter $j$ appears
such that  larger values amplify the $d_j$ effects. In this paper we will for
simplicity ignore these effects in accordance with arguments
that $j$ should take its smallest value \cite{Perez:2005fn, Vandersloot:2005kh}, as well
as questions as to the ambiguous nature of the critical scale at which the
corrections are appreciable. Further discussion can be found in \cite{kevthesis,Ashtekar:BBII,Newkplus1,Vandersloot:2006ws}.
Our results presented here will remain valid as long as the critical scale remains
below the scale at which the bounce occurs. In lieu of this, in our analysis
the matter energy density $\rho_M$ appearing in the effective Hamiltonian
\eqref{eqn:qm Hamiltonian} will not contain factors of $d_j$ and will
assume the classical form.

With these caveats in mind,
the effective Hamiltonian with lapse $N=\sqrt{p_1 p_2 p_3}$ is given by
\ba\label{eqn:qm Hamiltonian}
\Ham_{\rm eff}&=&
-\frac{1 }{\k \g^2}
\bigg\{
 \frac{\sin(\mubar_2c_2)\sin(\mubar_3c_3)}{\mubar_2 \mubar_3}\,p_2p_3\nn\\
&&\quad+\text{cyclic terms}
\bigg\}
+\,p_1p_2p_3\,\rho_M
\ea
and it becomes easy to see that in the limit of small $\mubar_I c_I$, the classical
Hamiltonian \eqref{Ham} is recovered.

Hamilton's equations proceed in the same fashion as the classical setup.
The equations for $dp_I/dt'$ and $dc_I/dt'$ give for instance
\be\label{eqn:qm dp/dt'}
\frac{dp_1}{dt'}=\frac{p_1\cos(\mubar_1c_1)}{\g}
\left\{{p_2}\frac{\sin(\mubar_2c_2)}{\mubar_2}+
{p_3}\frac{\sin(\mubar_3c_3)}{\mubar_3}\right\}
\ee
and
\ba\label{eqn:qm dc/dt'}
\frac{dc_1}{dt'}
&=&-\frac{1}{\g}
\left(\frac{3\sin(\mubar_1c_1)}{2\mubar_1}
-\frac{c_1\cos(\mubar_1c_1)}{2}\right)\nn\\
&&\quad\ \times
\left\{p_2\frac{\sin(\mubar_2c_2)}{\mubar_2}
+p_3\frac{\sin(\mubar_3c_3)}{\mubar_3}\right\}\nn\\
&&+\,\k\g\, p_2p_3\left(\rho_M+p_1\frac{\partial \rho_M}{\partial p_1}\right).
\ea
Hamilton's equations are now more complicated since
the discreteness parameters $\mubar_I$ depend on $p_I$, but let us define
$\mathcal{G}_I(t')$ as
\be \label{gI}
	\mathcal{G}_I(t') := p_I\frac{\sin(\mubar_Ic_I)}{\mubar_I},
\ee
which can be shown to satisfy
\be\label{gdot}
\frac{d \mathcal{G}_I}{dt'}
= \k \g \cos(\mubar_Ic_I)\, p_1p_2p_3\left( \r_M + p_I\frac{\partial \r_M}{\partial p_I}\right).
\ee
The vanishing of the Hamiltonian gives
\be\label{eqn:qm H=0}
	\k \g^2 \,p_1 p_2 p_3 \,\r_M = \mathcal{G}_1  \mathcal{G}_2 + \mathcal{G}_1 \mathcal{G}_3 + \mathcal{G}_2 \mathcal{G}_3.
\ee
Note that the classical limit is attained by  $\mubar_I c_I\rightarrow0$, where we have
$\sin(\mubar_Ic_I)/\mubar_I\rightarrow c_I$,
$\cos(\mubar_Ic_I)\rightarrow1$ and therefore
\eqnref{eqn:qm dp/dt'}, \eqnref{eqn:qm dc/dt'}, \eqnref{gdot} reduce to their
classical counterparts \eqnref{eqn:cl dp/dt'}, \eqnref{eqn:cl dc/dt'}, \eqnref{eqn:cl dpc/dt'}.

At this stage, the equations are too complicated to solve analytically
as was possible classically. To get a handle for the evolution, we can
consider as an example, the case of a massless scalar field with equation of state $w=+1$ whereby the equations
simplify considerably. With a massless scalar field, we have
\be
	\rho_M = \frac{P_{\phi}^2}{2 p_1 p_2 p_3},
\ee
where the momentum $P_{\phi}^2$ is a constant of motion.
Therefore, the term $\r_M + p_I\frac{\partial \r_M}{\partial p_I}$ in
\eqref{gdot} vanishes identically.
From this we find that
$\mathcal{G}_I$ are all constants in time. Equation \eqref{gI} then implies
that the triad components are all bounded as
\be
	p_I \ge \Big(|\mathcal{G}_I| \sqrt{\D}\Big)^{2/3},
\ee
implying that the classical singularity is never approached.
One can show from the equations of motion, that $\ddot{p}_I > 0$ when
$p_I = \big(|\mathcal{G}_I| \sqrt{\D}\big)^{2/3}$ implying that
the individual triad components $p_I$ bounce and hence the whole
universe must bounce.
With this bound
one can show that no curvature invariants blow up, and  the dynamics
is non-singular.\footnote{There is a special case when one of the $\mathcal{G}_I$
vanishes whence $p_I$ is not constrained above a minimum value. However,
this is the special case when $p_I$ is a constant which can be seen
from \eqref{eqn:qm dp/dt'}. Therefore the specific triad component
does not become zero and the evolution remains non-singular.} Note that this analysis holds for the vacuum case if $P_{\phi}=0$.
Therefore a bounce would occur generically for the vacuum case also.

Thus, at least with a massless scalar field, the bounce is robust under the
inclusion of anisotropies. Next we can monitor the behavior
of the anisotropic shear term through the bounce for this example.
If we look at the shear parameter $\Sigma$ we showed that classically it remains
constant throughout the evolution. With the loop quantum modifications
we can monitor its behavior by using its definition
\be
	 \Sigma^2 :=   \frac{a^6}{18} \bigg( (H_1-H_2)^2 + (H_2-H_3)^2
	+ (H_3-H_1)^2  \bigg),
\ee
which in the classical case was a constant because of equation \eqref{HI}.
In the LQC case, let us assume we start with a nearly classical contracting universe
with each $\mubar_Ic_I \ll 1$. In this limit from \eqref{gI} we find
\be
	\mathcal{G}_I \approx p_I c_I.
\ee
Since $\mathcal{G}_I$ are constant for the massless scalar field, we can identify
them with the classical constants $\alpha_{IJ}$ from
\eqref{aIJ} as $\mathcal{G}_I - \mathcal{G}_J  = \g \Vz \alpha_{IJ}$.
The shear factor $\Sigma$ is then given initially as
\be
	\Sigma^2(\text{pre bounce}) \approx \frac{1}{18 \g \Vz} \bigg( ( \mathcal{G}_1 - \mathcal{G}_2)^2
 +\text{cyclic terms}  \bigg).
\ee
After the bounce, the constancy of $\mathcal{G}_I$ and  equation
\eqref{gI} imply that $\mubar_I c_I$ become small as $p_I$ grow
and hence  classical behavior is recovered at late times. We can apply
the same argument to conclude that the late time behavior of $\Sigma$
approaches the pre-bounce value in terms of the same constants $\mathcal{G}_I$:
\be
	\Sigma^2(\text{post bounce})  = \Sigma^2(\text{pre bounce})
\ee
and hence the shear factor $\Sigma$ is conserved before and after the bounce.
Using equations \eqref{eqn:qm dp/dt'} and \eqref{gI}, it is not too difficult
to show that the shear term remains finite through the bounce indicating
that the anisotropies do not blow up.

The similar conclusions can be obtained for the generic cases with the inclusion of arbitrary matter with $w<+1$. The details of the effective loop quantum dynamics with generic perfect fluids are investigated in \appref{app2}. The detailed analysis shows that individual $p_I$ in different directions can  bounce at slightly different moments. Moreover, there is competition between the matter energy density $\r_M$ and the
\emph{directional density} $\vr_I$ (defined in \eqnref{eqn:directional density}, which is associated with the classical anisotropic shear) to be dominant when the quantum corrections start to become significant. As a result, depending on how energetic the matter content is (compared to the degree of anisotropies), the bounce can take place either in the ``Kasner phase'' or in the ``isotropized phase'' (or in the ``transition phase'' in between). If quantum corrections take effect in the isotropized phase, the bounce happens around the moment when $\r_M$ approaches $.82\r_{\rm Pl}$, giving similar results as in the isotropic model. On the other hand, if the bounce occurs in the Kasner phase, the individual triad components bounce
when
$\vr_I$ approach $.86\r_{\rm Pl}$.

In the case that the big bounce occurs in the isotropized phase, the Kasner fashion of the classical solution is smeared by the quantum effect and thus the information of anisotropies is blurred. Therefore, on the other side of the bounce, the classical anisotropic shear is changed. This explains why $\Sigma^2(\text{post bounce})\neq\Sigma^2(\text{pre bounce})$ in general. Only in the case when the big bounce takes place in the Kasner phase do we have $\Sigma^2(\text{post bounce})\approx\Sigma^2(\text{pre bounce})$.\footnote{In the cases of vacuum and with only massless scalar field, the classical solution remains in Kasner phase throughout the evolution (the universe is not isotropized by matter). Thus, quantum corrections always take effect in the Kasner phase and this is why we have $\Sigma^2(\text{post bounce})=\Sigma^2(\text{pre bounce})$ exactly for these two special cases.}

In the next section we explicitly show the results mentioned here by numerically solving the equations of motion.

\section{Numerical Results}
Owing to the complexity of the effective equations of motion, we can not
go further in the analytical analysis for more general forms of matter. In this section we will present some
numerical simulations of the equations of motion by including other forms of matter.
In particular we will focus on whether a bounce occurs for an initially contracting
universe and how the shear term $\Sigma$ behaves through the bounce. Additionally
we will show that the universe isotropizes after the bounce with the inclusion of $w < +1 $ matter.

The differential equations for the time evolution of $p_I(t')$ and $c_I(t')$ are
given in
\eqref{eqn:qm dp/dt'} and \eqref{eqn:qm dc/dt'}. Numerically we solve
the equations given a lapse $N$ equal to one and plot as a function of cosmic time $t$.
The initial conditions are chosen consisting of a collapsing semi-classical universe such
that $\dot{a}< 0$ and $\mubar_I c_I \ll 1$. The shear term $\Sigma$ is monitored
using the classical formula \eqref{shear}.

The first example we consider is the vacuum case. The analysis can be understood analytically
as a special case of the massless scalar field studied in the previous section. The same
analysis holds when $P_{\phi} = 0$ which gives the vacuum case. Therefore,
for a vacuum case a bounce is generic and the shear term $\Sigma^2$ is
conserved before and after the bounce.

\begin{figure*}
\centering
	\subfigure[]{
	\label{avacuum}
	\includegraphics[width=8cm, keepaspectratio]
        {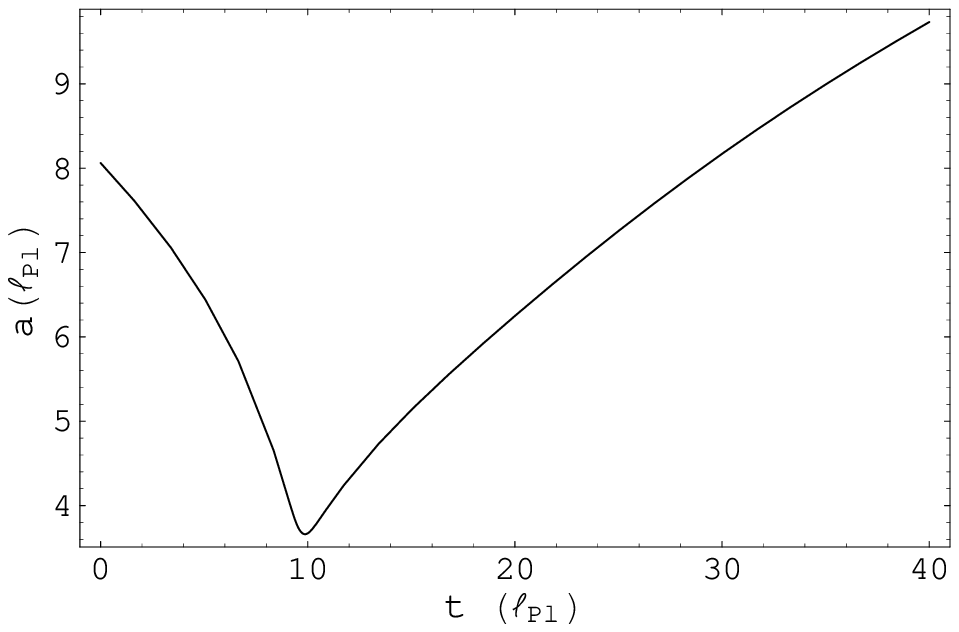}
	}
	\hspace{.3in}
	\subfigure[]{
	\label{aivacuum}
	\includegraphics[width=8cm, keepaspectratio]
        {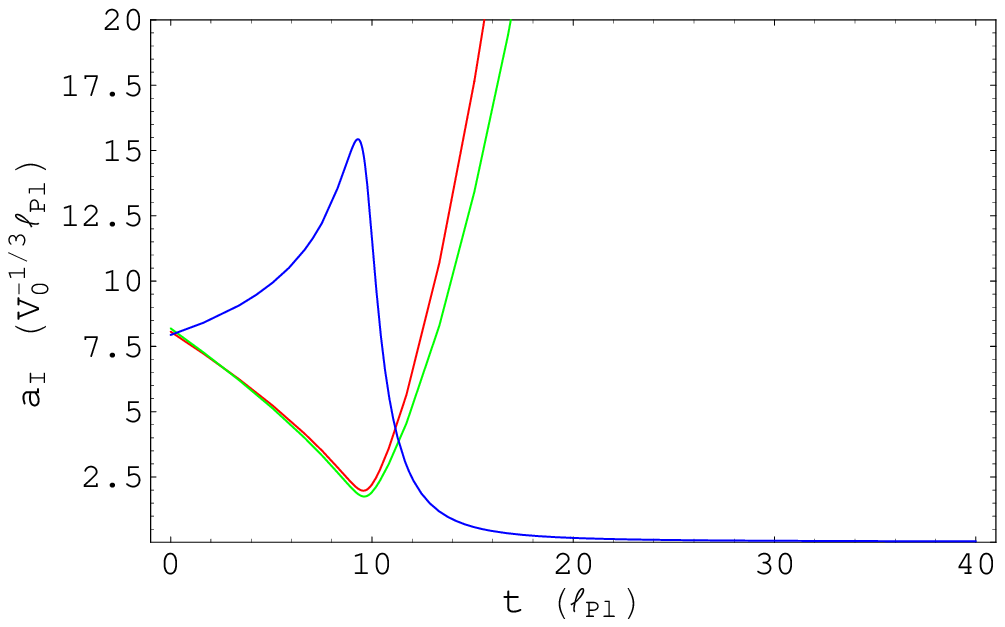}
	}
\caption{Mean scale factor $a(t)$ and directional scale factors $a_I(t)$ for
vacuum $\r_M =0$ case. The initial conditions are chosen corresponding to
an initially contracting universe.}
\label{asvacuum}
\end{figure*}

\begin{figure}[ht]
\begin{center}
\includegraphics[width=7.5cm, keepaspectratio]
        {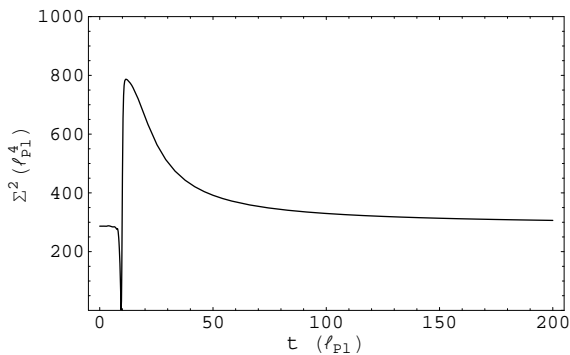}
\end{center}
\caption{Shear term for vacuum case. Classically, the value is a constant, though
quantum mechanically it is not constant near the bounce. The post-bounce value
approaches the pre-bounce value, and nowhere does it blow up.}
\label{ssqvacuum}
\end{figure}

The mean scale factor $a(t)$, directional scale factors $a_I(t)$, and
shear term $\Sigma^2(t)$ are plotted in Figures \ref{asvacuum}
and \ref{ssqvacuum} respectively, for a representative
numerical simulation consisting of an initially contracting universe which
in the vacuum Kasner case requires two contracting directions and one
expanding. After the bounce, the universe consists of a Kasner like
expanding phase with growth in two directions and contraction in the other.
Throughout the evolution, the shear term remains finite with the behavior during
the bounce showing no obvious pattern (from numerical simulations, the exact behavior
can depend on initial conditions and choice of matter). However, it is clear
that the post-bounce value tends to the pre-bounce value as
shown by the analytical arguments of the previous section.

\begin{figure*}
\centering
	\subfigure[]{
	\label{astiff}
	\includegraphics[width=8cm, keepaspectratio]
        {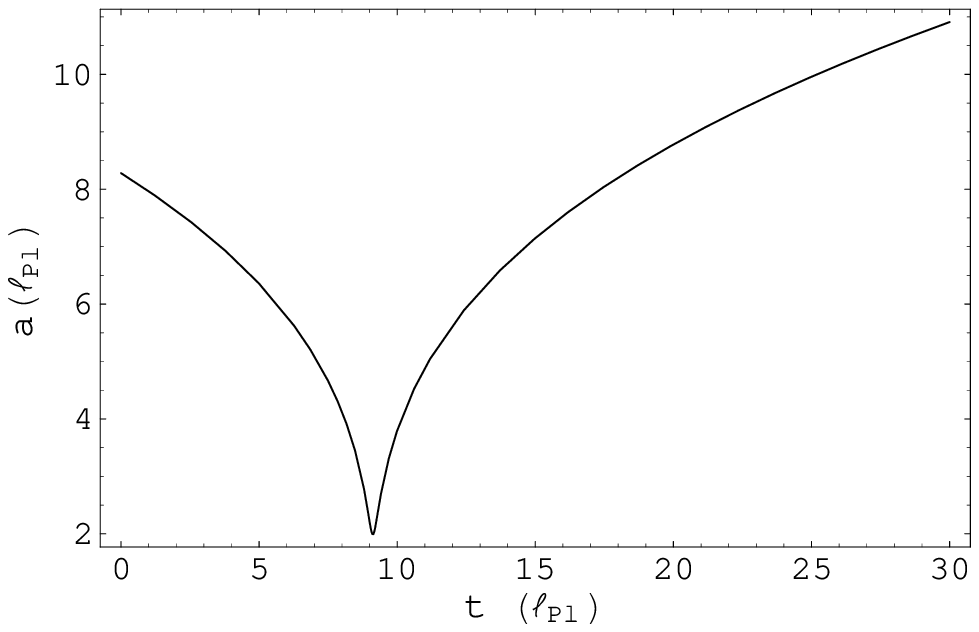}
	}
	\hspace{.3in}
	\subfigure[]{
	\label{aistiff}
	\includegraphics[width=8cm, keepaspectratio]
        {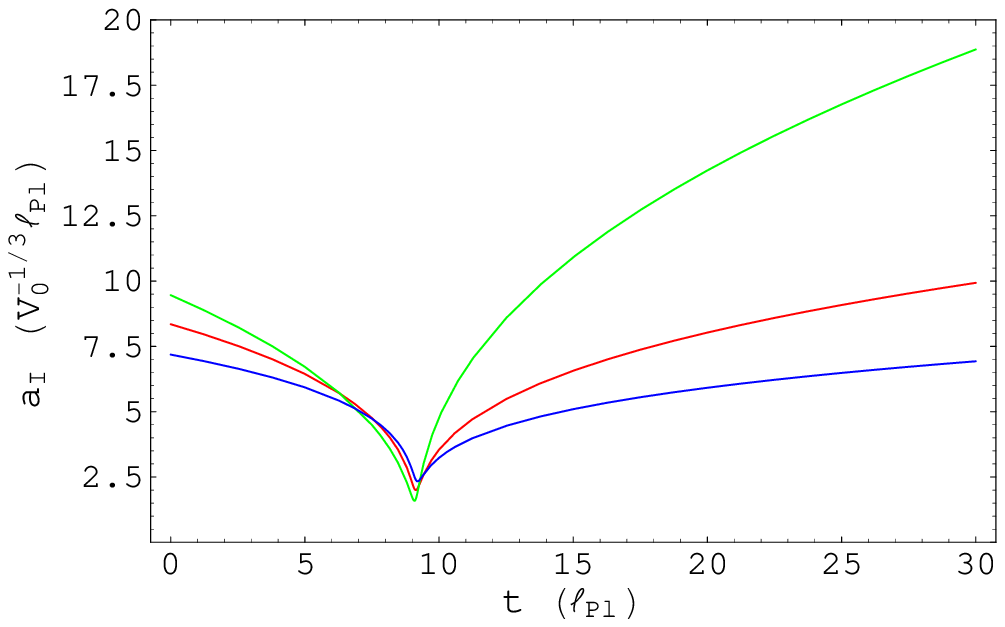}
	}
\caption{Mean scale factor $a(t)$ and directional scale factors $a_I(t)$ for
a massless scalar field with momentum $P_{\phi} = 10 l_p^2$. Initially, all directions are contracting and a bounce occurs leading to expansion in all directions.}
 \label{asstiff}
\end{figure*}

\begin{figure}[ht]
\begin{center}
\includegraphics[width=7.5cm, keepaspectratio]
        {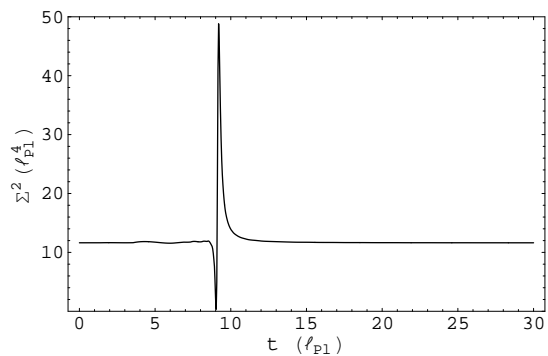}
\end{center}
\caption{Shear term for massless scalar field. As in the vacuum case, the pre-bounce value
is conserved after the bounce.}
\label{ssqstiff}
\end{figure}

The inclusion of a massless scalar field does not change the results of
the vacuum case significantly. With matter, the initial contracting phase can
be either one where all directions contract, or a Kasner like contraction as in the
vacuum case. In Figures \ref{asstiff}
and \ref{ssqstiff}, the scale factors and shear term are plotted for the case
where all three directions initially are contracting. Again, the shear term is
finite through the evolution and the post-bounce value approaches the pre-bounce
value, as expected from the analytical analysis.

\begin{figure*}
\centering
	\subfigure[]{
	\label{arad}
	\includegraphics[width=8cm, keepaspectratio]
        {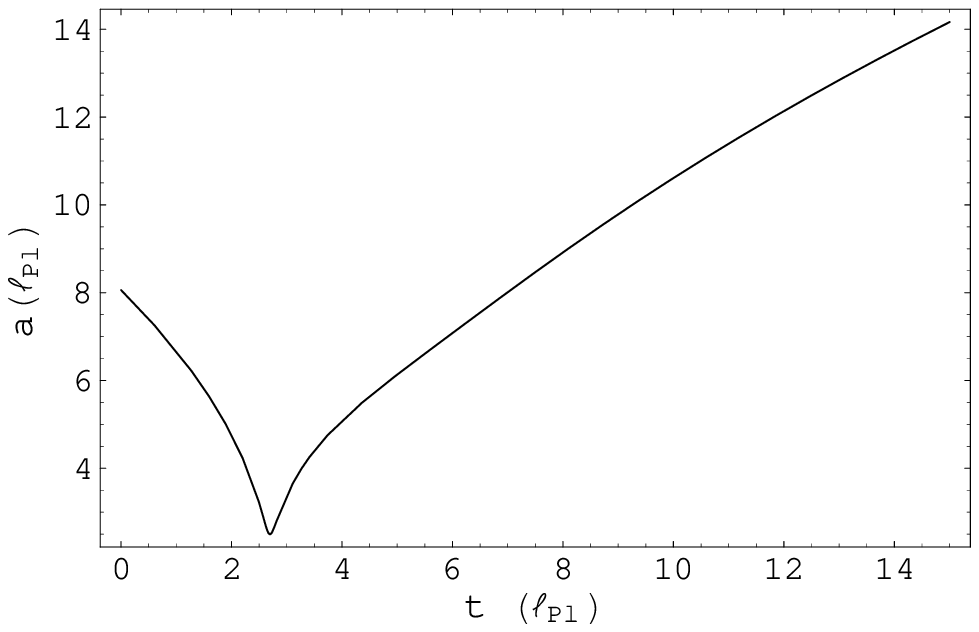}
	}
	\hspace{.3in}
	\subfigure[]{
	\label{airad}
	\includegraphics[width=8cm, keepaspectratio]
        {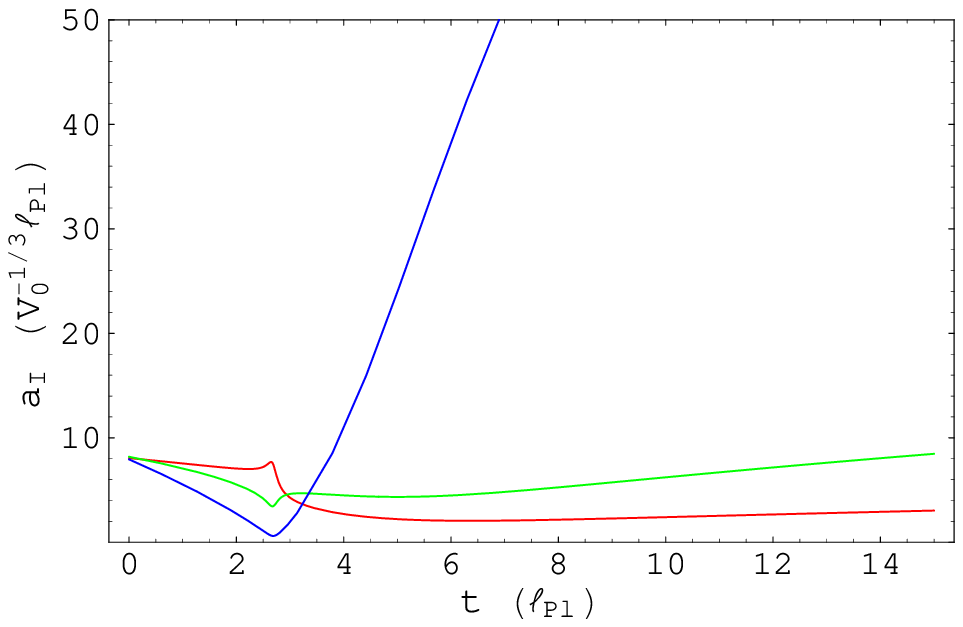}
	}
\caption{Mean scale factor $a(t)$ and directional scale factors $a_I(t)$ for
radiation.}
\label{asrad}
\end{figure*}

\begin{figure}[ht]
\begin{center}
\includegraphics[width=7.5cm, keepaspectratio]
        {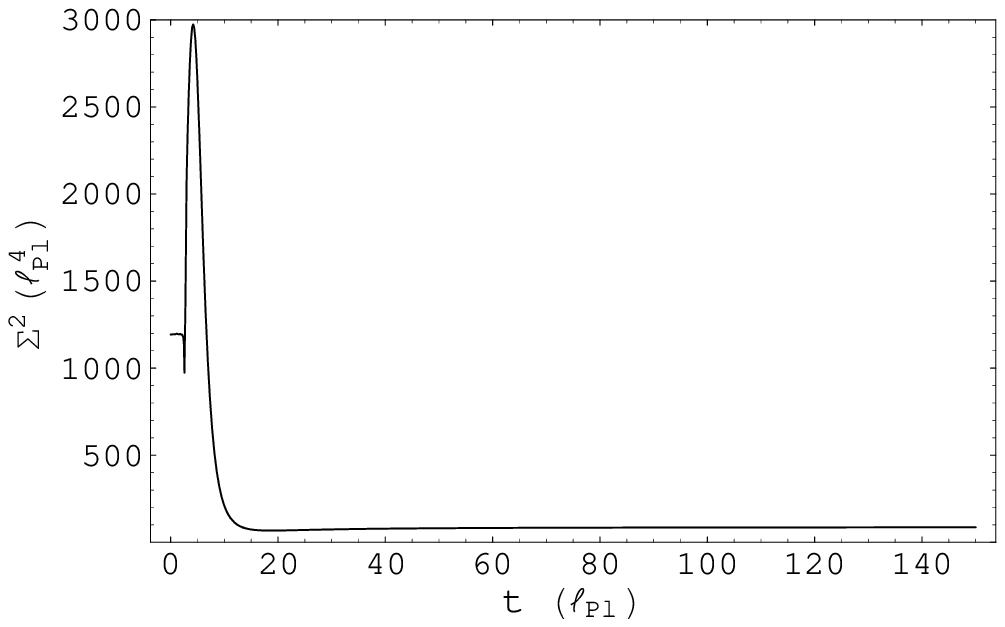}
\end{center}
\caption{Shear term for radiation. The post-bounce value does not equal the pre-bounce
values although the value is finite through the entire evolution.}
\label{ssqrad}
\end{figure}

As an example of the inclusion of other forms of matter, we consider a radiation ($w=1/3$)
field with energy density
\be
	\r_M = \frac{\tilde{A}}{a^4}
\ee
with $\tilde{A}$ some constant. The numerical behavior of the scale factor appears
in Figure \ref{asrad}, where again the bounce occurs. Various initial conditions
were examined and in all cases the bounce occurred providing evidence that
it is a general feature of the effective Hamiltonian we study in this paper. In contrast
to the massless scalar field case, the shear term is not conserved after the bounce.
This is shown in Figure \ref{ssqrad}, and for the particular initial conditions
chosen, actually decreases in the post-bounce regime. It must be stated that
the decrease is not a generic feature and is sensitive to the initial conditions.
However, as in the previous cases, the shear term is finite through the entire
evolution.

\begin{figure}[ht]
\begin{center}
\includegraphics[width=7cm, keepaspectratio]
        {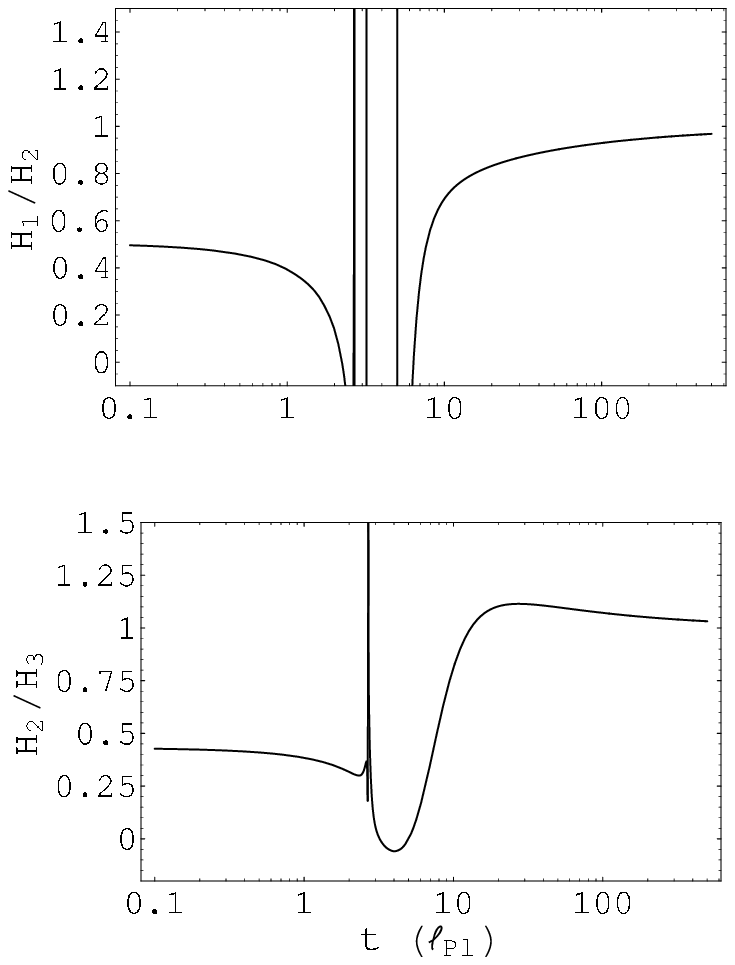}
\end{center}
\caption{Ratios of directional Hubble rates for radiation. At late times,
the ratios approach unitary indicative of equal expansion rates in
all directions as the universe isotropizes.}
\label{hubsrad}
\end{figure}

Since the equation of state for the radiation field $w=1/3 < +1$, the post-bounce
regime leads to an isotropization of the universe. This is borne out in
the numerical simulations in Figure \ref{hubsrad}. There are plotted
ratios of the directional Hubble ratios. The ratios approach unity
in the post-bounce epoch indicating identical expansion rates in
the three directions.

\section{Discussion}
Let us restate the main results presented. We have consider the anisotropic Bianchi I
model with loop quantum corrections to the classical equations of motion.
We have shown that a bounce occurs under rather generic conditions in
a collapsing universe. The anisotropic shear term $\Sigma^2 / a^6$, which
classically grows during collapse and blows up at the singularity, remains finite
through the bounce. After the bounce, the universe behaves more and more
classically and can isotropize at later times. This is thus evidence that
the bouncing scenario of the isotropic models of loop quantum cosmology is
robust when the symmetries of the isotropic model are relaxed.

We must reiterate the caveats that has gone into this analysis. First,
we have used entirely the effective classical equations of motion
determined from the effective Hamiltonian \eqref{eqn:qm Hamiltonian}.
This effective Hamiltonian is  motivated from the construction of
the quantum Hamiltonian operator in a somewhat heuristic fashion. In principle, however, there can
be additional modifications arising from the quantum theory. To properly
justify the analysis here, more work is required to analyze the quantum dynamics
by constructing semi-classical states and evolving them with the quantum difference equation.
The accuracy of the effective equations of motion has been established in
the isotropic case with a massless scalar field, and thus we expect that they
should give an accurate representation of the dominant corrections arising
in the quantum dynamics.

In addition, we have ignored the inverse triad effects arising in loop quantum
cosmology. One might ask if we do include these effects, does the
bounce picture still hold. Yet, we can answer this by noting that the inverse triad
modifications tend to suppress the matter energy density. Thus if the suppression is
large, the universe behaves more like a vacuum Bianchi I model. Our analysis indicates
that the bounce still occurs in this case. Therefore, the inverse triad effects are not
expected to remove the bounce, and only change the quantitative behavior.

We have not considered in this paper additional anisotropic Bianchi models.
It would be interesting to extend the analysis to these models. In particular
the results in principle could  be extended to the Bianchi IX model. An interesting
question would be whether the quantum effects tame the classical chaos
in the model and again whether a bounce is predicted. Additionally, since the
Bianchi IX model has been conjectured to be of relevance to singularities
in general, the analysis may provide hints as to what role loop quantum effects
play with regards to general singularities.

\acknowledgements
The authors would like to thank Abhay Ashtekar, Martin Bojowald, Golam Hossain, Roy Maartens, Tomasz Pawlowski and Parampreet Singh for useful discussions.
This work was supported in part by
the NSF grant PHY-0456913, the Eberly research funds of Penn State,
and the Marie Curie grant MIF1-CT-2006-022239.

\appendix

\section{Details of the classical solutions}\label{app1}
In this appendix we discuss in more detail the solutions to the classical equations of
motion. We show explicitly how the universe near the singularity is dominated by
Kasner like dynamics and how the universe can
isotropize far away from the singularity for matter with $w<1$.

Let us start with the relation \eqnref{eqn:cl key relation}, which, in addition  to \eqnref{aIJ}, implies
\be\label{eqn:K+f}
p_Ic_I=\k\g\hbar\left[\K_I+f(t')\right]
\ee
with the dimensionless constants of motion $\K_I$. Note that
\be
\k\hbar(\K_I-\K_J)=\Vz\a_{IJ}.
\ee
We assume that the matter density is in the form
\be
\r_M=A\,(p_1p_2p_3)^{-(1+w)/2}
\ee
with $A$ a constant and $w$ the state parameter.

The Hamiltonian constraint $\Ham=0$ with $\Ham$ given by \eqnref{Ham} then yields
\ba
&&3f(t')^2+2\left(\K_1+\K_2+\K_3\right)f(t')\\
&&\quad+\K_2\K_3+\K_1\K_3+\K_1\K_2
=(\k\hbar^2)^{-1}A\left(p_1p_2p_3\right)^\frac{1-w}{2},\nn
\ea
which gives the time-independent part:\footnote{We can always absorb an arbitrary constant to $f(t')$ and thus change the dichotomy between the time-independent and time-dependent parts. However, we pick the particular choice as in \eqnref{eqn:KK} and \eqnref{eqn:f(t')} in order to relate $\K_I$ to the standard parameters used in the Kasner solutions.}
\be\label{eqn:KK}
\K_2\K_3+\K_1\K_3+\K_1\K_2=0
\ee
and the time-dependent part:
\ba\label{eqn:f(t')}
&&f(t')=-\frac{\K_1+\K_2+\K_3}{3}\\
&&\quad\pm\frac{1}{3}\left[(\K_1+\K_2+\K_3)^2
+\frac{3A\left(p_1p_2p_3\right)^\frac{1-w}{2}}
{\k\hbar^2}\right]^{1/2}.\nn
\ea
We can scale the constants $\K_I=\K\k_I$ such that \eqnref{eqn:KK} gives
\be\label{eqn:Kasner condition}
\k_1+\k_2+\k_3=1,\qquad
\k_1^2+\k_2^2+\k_3^2=1,
\ee
which coincide with the ``Kasner condition'' satisfied by the parameters $\k_I$ used for the vacuum Bianchi I solutions (Kasner solutions). [For the Kasner solutions, apart from the trivial solution (Minkowski spacetime), two of $\k_I$ must be positive while the other negative, giving the universe expanding in two direction and contracting in the other (or the other way around if $\K<0$).]
Furthermore, note that
the opposite choice of the sign $\pm$ in \eqnref{eqn:f(t')} amounts to the changes: $f(t')\rightarrow -f(t')$ and $\K_I\rightarrow -\K_I$ simultaneously, which correspond to the time reversal. Therefore, without losing generality, we can stick with positive sign for $A>0$; for $A<0$, on the other hand,
$\pm$ flips sign when the part inside the square bracket of \eqnref{eqn:f(t')} approaches zero.\footnote{For $A<0$, $\pm$ changes signs and thus the solutions of \eqnref{eqn:de of p} may encounter recollapse. This is the case for the model with negative cosmological constant $\L<0$.} We assume $A>0$ in this paper for ordinary matter with positive energy.
With \eqnref{eqn:K+f} and \eqnref{eqn:f(t')}, the equation of motion \eqnref{eqn:cl dp/dt'} gives
\ba\label{eqn:de of p}
&&\frac{1}{p_1}\frac{dp_1}{dt'}=\frac{1}{\g}\left(c_2p_2+c_3p_3\right)
=\k\hbar\left[\K_2+\K_3+2f(t')\right]\nn\\
&=&\k\hbar\bigg\{\frac{\K_2+\K_3-2\K_1}{3}\\
&&\quad+\frac{2}{3}\left[(\K_1+\K_2+\K_3)^2
+\frac{3A\left(p_1p_2p_3\right)^\frac{1-w}{2}}
{\k\hbar^2}\right]^{1/2}\bigg\}.\nn
\ea

For $w<1$, the second term in the square bracket of \eqnref{eqn:de of p} is negligible when the solution approaches the big bang singularity; on the opposite side, it becomes dominant in the large universe limit.\footnote{For $w=1$, the second term becomes constant and so does $f(t')$; as a result, the solutions have qualitatively different features. This is the case with massless scalar field. In particular, the scalar field can be treated as ``internal time'' and the Kasner condition \eqnref{eqn:Kasner condition} is modified such that ``Kasner-unlike'' solutions (namely, expanding/contracting in \emph{all} three directions) are also allowed \cite{Chiou:2006qq}; plus, the anisotropy persists in the expanding phase if no other matter content is included.} Therefore, in the vicinity of the singularity, \eqnref{eqn:de of p} is approximated as
\be\label{eqn:cl approx sol}
\frac{1}{p_1}\frac{dp_1}{dt'}\approx
{\k\hbar}\left(\K_2+\K_3\right),
\ee
yielding the solutions very close to Kasner solutions, which are highly anisotropic.
On the other hand, in the large universe limit, \eqnref{eqn:de of p} have the asymptotic behavior
\be\label{eqn:cl asymp sol}
\frac{1}{p_1}\frac{dp_1}{dt'}\equiv
\frac{(p_1p_2p_3)^{1/2}}{p_1}\frac{dp_1}{dt}\approx
2\sqrt{\frac{\k A}{3}}\,\left(p_1p_2p_3\right)^\frac{1-w}{4},
\ee
giving the asymptotic solution
\be\label{eqn:cl asymp sol2}
p_I(t)\propto\left\{
\begin{array}{ll}
 t^{\frac{4}{3(1+w)}} &\qquad w\neq -1,\\
e^{2t\sqrt{\frac{\k A}{3}}} &\qquad w=-1.
\end{array}
\right.
\ee
%with
%\be
%A_1A_2A_3=\left[\frac{2}{1-w}\right]^\frac{4}{1-w}
%\left(\frac{3A}{\k\hbar^2}\right)^{-\frac{2}{1-w}}.
%\ee
As the universe approaches the asymptotic region, the three directions are all expanding with the same rate; that is, with the inclusion of matter with equation of state $w<1$, the universe isotropizes in the expanding phase.
When the contribution from matter sector is negligible and the evolution is essentially the same as the Kasner solution as given by \eqnref{eqn:cl approx sol}, we call it ``Kasner phase''. On the opposite, when the matter sector dominates and the universe is isotropized as given by \eqnref{eqn:cl asymp sol}, we call it ``isotropized phase''. The situation in between is called ``transition phase''.

As anisotropy is concerned, we study the Hubble ratios, which give the asymptotic behaviors:
%\be
%\Upsilon^I_J:=\frac{H_I}{H_J}\equiv
%\frac{\dot{a}_I/a_I}{\dot{a}_J/a_J}.
%\ee
\be
\frac{H_I}{H_J}\equiv
\frac{\dot{a}_I/a_I}{\dot{a}_J/a_J}\approx\left\{
\begin{array}{cll}
\k_I/\k_J & \quad\text{for} & a\rightarrow 0,\\
1 & \quad\text{for} & a\rightarrow \infty,
\end{array}
\right.
\ee
which are implied by equations \eqnref{eqn:cl approx sol} and \eqnref{eqn:cl asymp sol}.
These ratios approach unity in the isotropized phase and approach fixed constants in the Kasner phase.\footnote{Note that, however, in the special case of vacuum (A=0; i.e Kasner model) or of scalar matter ($w=1$), the Hubble ratios $H_I/H_I$ are constant throughout the entire evolution. The classical solution does not isotropize and remains in the Kasner phase.}

\section{Details of the effective loop quantum solutions}\label{app2}
In order to affirm the assertions in \secref{effective theory} for the generic cases with arbitrary matter, this section deals with the detailed analysis for the effective loop quantum solutions.
The effective dynamics with quantum corrections is governed by \eqnref{eqn:qm dp/dt'}, \eqnref{eqn:qm dc/dt'} and \eqnref{eqn:qm H=0}, which are complicated to solve analytically for the generic case. However, we can still study two extreme cases: $a\rightarrow\infty$ and $a\rightarrow 0$.

In the limit $a\rightarrow\infty$, if $\mubar_Ic_I\rightarrow0$, the Hamilton's equations plus the Hamiltonian constraint simply reduce to their classical counterparts and the effective solutions are virtually the same as the classical one. We do not know  \textit{a priori} whether $\mubar_Ic_I\rightarrow0$ for large $a$ and whether the quantum effect does not spoil the semi-classicality. But if the classical solutions without quantum corrections leads to small value of $\mubar_Ic_I$ for large universe, we can conclude that the effective dynamics admits the solutions with semi-classicality for large $a$. To check this, consider the classical solutions, which have
\ba
p_Ic_I&=&\k\g\hbar\left[\K_I+f(t')\right]\approx\k\g\hbar f(t')\nn\\
&\approx& \g\sqrt{\frac{\k A}{3\hbar}}\,\left(p_1p_2p_3\right)^{\frac{1-w}{4}}
\ea
for $a\rightarrow\infty$ by \eqnref{eqn:K+f}.
We then have, by \eqnref{eqn:cl asymp sol2},
\be
c_I\propto a^{-\frac{(1+3w)}{2}},
\qquad
\mubar_I c_I\propto
a^{-\frac{3(1+w)}{2}}.
\ee
This shows that the quantity $\mubar_Ic_I$ of classical soultions is decreasing to zero in the large universe limit for $w>-1$.
Therefore, for $-1<w<1$, at large universe, the semi-classicality can be retained and the loop quantum corrections are indeed negligible for those solutions which are semi-classical at late times. The evolution simply follows the classical trajectory for the large universe.\footnote{This is also true for the special cases of vacuum and of scalar matter with $w=1$, in both of which $p_Ic_I=\text{constant}$ and thus $\mubar_Ic_I\propto p_I^{-3/2}\rightarrow0$ classically. For the case of the cosmological constant ($w=-1$), the cosmological constant $\L$ has to be very small to admit the semi-classicality.}

Now, let us study the other extreme as the universe approaches the classical singularity ($a \rightarrow 0$). In the backward evolution, before the singularity is reached, the quantum corrections will take effect and the big bounce is expected to take place. Therefore, at some point, $\cos(\mubar_I c_I)$ vanishes and flips sign. Assuming that $\cos(\mubar_I c_I)$
in different directions flip sign at only slightly different moments, we then have $\cos(\mubar_I c_I)\approx \cos(\mubar_J c_J)$ ($\rightarrow0$) in the vicinity of the big bounce.
With this approximation, close to the epoch of the big bounce, \eqnref{gdot} yields\footnote{The assumption that $\cos(\mubar_I c_I)$ flip signs at only slightly different moments could be wrong in general. However, note that even if $\cos(\mubar_I c_I)\approx\cos(\mubar_J c_J)$ does not hold very well, the vanishing of \eqnref{eqn:qm key relation} is still a good approximation near the bounce, since $(p_1p_2p_3)^{(1-w)/2}=a^{3(1-w)}\rightarrow0$ for $w<1$ when $a$ is small enough. (But the condition $\cos(\mubar_I c_I)\approx\cos(\mubar_J c_J)$ makes the approximation \eqnref{eqn:qm key relation} even more accurate.) In the special case for $w=1$, \eqnref{eqn:qm key relation} vanishes \emph{exactly} simply because the factor $(1-w)/2$ is zero.}
\ba\label{eqn:qm key relation}
&&\frac{d}{dt'}\left(\mathcal{G}_I-\mathcal{G}_J\right)\\
&=&\frac{1-w}{2}\k\g
\left[\cos(\mubar_I c_I)- \cos(\mubar_J c_J)\right]
\left(p_1p_2p_3\right)^\frac{1-w}{2}
\approx 0.\nn
\ea
Thereby, we can write
\be\label{eqn:qm G_I}
\mathcal{G}_I\equiv p_I\frac{\sin(\mubar_I c_I)}{\mubar_I}
\approx \k\g\hbar\left[\K_I+f(t')\right]
\ee
in accordance with the classical counterpart \eqnref{eqn:K+f};
consequently the Hamiltonian constraint \eqnref{eqn:qm H=0} gives the same $f(t')$ as given by \eqnref{eqn:f(t')}.
To proceed further, in the following, we consider three cases separately:
quantum corrections take effect
(i) in the Kasner phase;
(ii) in the isotropized phase; and
(iii) in the transition phase.

In Case (i), the first term in the square bracket dominates over the second term in \eqnref{eqn:f(t')}; thereby,
\ba
f(t')&\approx&-\frac{\K}{3}+\frac{\K}{3}
\left[1+\frac{3A\left(p_1p_2p_3\right)^\frac{1-w}{2}}{2\K^2\k\hbar^2}
+\cdots\right]\nn\\
&\approx&
\frac{ A}
{2\K\k\hbar^2}\left(p_1p_2p_3\right)^\frac{1-w}{2}.
\ea
It then follows from \eqnref{eqn:qm G_I} that
\ba\label{eqn:qm sin}
\sin(\mubar_I c_I)
&\approx&
\k\g\hbar\frac{\D^{1/2}}{p_I^{3/2}}
\left\{\K_I
+\frac{A}
{2\K\k\hbar^2}\left(p_1p_2p_3\right)^\frac{1-w}{2}
\right\}\nn\\
&\approx&\k\g\hbar\frac{\D^{1/2}\K_I}{p_I^{3/2}},
\ea
provided
\be\label{eqn:case i condition}
\K_I\gg\frac{A}{\K\k\hbar^2}\left(p_1p_2p_3\right)^\frac{1-w}{2}.
\ee

Taking \eqnref{eqn:qm G_I} into \eqnref{eqn:qm dp/dt'} and expressing $\cos x=\pm(1-\sin^2\!x)^{1/2}$ with the help of \eqnref{eqn:qm sin}, we get for instance
\ba\label{eqn:bouncing p_I}
\frac{1}{p_1}\frac{dp_1}{dt'}&\approx&
\pm\k\hbar\left(\K_2+\K_3+\cdots\right)
\nn\\
&&\qquad\times
\left[1-\frac{(\k\g\hbar)^2\D}{p_1^3}
\left(\K_I
+\cdots
\right)^2
\right]^{1/2}\nn\\
&\approx&\pm\k\hbar\left(\K_2+\K_3\right)
\left[1-\frac{p^3_{1,{\rm crit}}}{p_1^3}\right]^{1/2},
\ea
where the critical value $p_{I,{\rm crit}}$ is given by the Planck length square $\Pl^2$ times a numerical factor:
\ba\label{eqn:p_I crit}
p_{I,{\rm crit}}&:=&\abs{\K_I}^{2/3}
\left(\k\g\hbar\right)^{2/3}\D^{1/3}\nn\\
&\approx&
19.01\g\abs{\K_I}^{2/3}\Pl^2.
\ea

Therefore, the big bang singularity is replaced by the bounces whenever each of $p_I$ approaches its critical value $p_{I,{\rm crit}}$. The bounces occur up to three times, once in each diagonal direction. If we define the \emph{directional density} $\vr_I$ in the $I$-direction as:
\be\label{eqn:directional density}
\vr_I:=\frac{\k\hbar^2\K_I^2}{3p_I^3},
\ee
the above statement can be rephrased to say: The big bounces take place whenever each of the directional densities reaches the critical value
\be\label{eqn:rho_Pl}
\vr_{\rm crit}=3(\k\g^2\D)^{-1}
\approx .82\r_{\rm Pl}.
\ee
Note that $\vr_I$ have the same dimension as $\r_M$ and moreover we have
\be
\k^{-1}\frac{\bar{\Sigma}^2}{a^6}=\frac{1}{3}
\left[\frac{p_1^3\vr_1+p_2^3\vr_2+p_3^3\vr_3}{p^3}\right]
\ee
with
\be
\bar{\Sigma}^2:=\frac{1}{18}\left(\a_{12}^2+\a_{23}^2+\a_{31}^2\right)
\ee
being identical to the shear factor $\Sigma^2$ of the classical solution. This suggests that $\vr_I$ can be roughly interpreted as the ``energy density carried from the classical anisotropic shear portioned for the $I$-direction'' (and thus the name).

To meet the condition \eqnref{eqn:case i condition} of Case (i), we take the critical values $p_{I,{\rm crit}}$ into \eqnref{eqn:case i condition} and find the criterion for Case (i) to be
\be\label{eqn:case i criterion}
A\ll\frac{\abs{\k_I}\g^{w-1}}{\abs{\k_1\k_2\k_3}^{(1-w)/3}}\,
\K^{1+w}\k^w\hbar^{1+w}\D^\frac{w-1}{2}.
\ee

Next, in Case (ii), on the other hand, the second term in the square bracket of \eqnref{eqn:f(t')} dominates and we have
\be
f(t')\approx\sqrt{\frac{A}{3\k\hbar^2}}
\left(p_1p_2p_3\right)^\frac{1-w}{4}.
\ee
It then follows from \eqnref{eqn:qm G_I} that
\be\label{eqn:qm sin2}
\sin(\mubar_I c_I)\approx
\k\g\hbar\frac{\D^{1/2}}{p_I^{3/2}}
\sqrt{\frac{A}{3\k\hbar^2}}
\left(p_1p_2p_3\right)^\frac{1-w}{4},
\ee
provided
\be\label{eqn:case ii condition}
\K_I\gg\frac{A}{\k\hbar^2}\left(p_1p_2p_3\right)^\frac{1-w}{2}.
\ee
With the help of \eqnref{eqn:qm sin} again, \eqnref{eqn:qm dp/dt'} gives
\ba
\frac{1}{p_1}\frac{dp_1}{dt'}&\approx&
\pm2\k\hbar\sqrt{\frac{A}{3\k\hbar^2}}
\left(p_1p_2p_3\right)^\frac{1-w}{4}
\\
&&\qquad\times
\left[1-\frac{(\k\g\hbar)^2\D}{p_1^3}
\frac{A}{3\k\hbar^2}\left(p_1p_2p_3\right)^\frac{1-w}{2}
\right]^{1/2}.\nn
\ea

The big bang singularity is again replaced by the bounces and the bouncing points of $p_I$ are roughly equal in all three direction; i.e., each $p_I$ is bounced when $p_I\approx p_{\rm crit}$ with
\be\label{eqn:p crit}
p_{\rm crit}:=\left[\frac{\k\g^2\D A}{3}\right]^\frac{2}{3(1+w)}.
\ee
For more generic cases (with multiple matters), this means that the bounces take place near the point when the (total) matter density approaches its critical value
\be
\r_{\rm crit}\equiv A p_{\rm crit}^{-3(1+w)/2}=3(\k\g^2\D)^{-1}
\approx .86\r_{\rm Pl},
\ee
begin the same as the critical value given in \eqnref{eqn:rho_Pl}.

To meet the condition \eqnref{eqn:case ii condition} of Case (ii), we take the critical value $p_{\rm crit}$ into \eqnref{eqn:case ii condition} and find the criterion for Case (ii) to be
\be\label{eqn:case ii criterion}
A\ll\g^{w-1}
\K_I^{1+w}\k^w\hbar^{1+w}\D^\frac{w-1}{2},
\ee
which is exactly the opposite of the criterion \eqnref{eqn:case i criterion} if we have $\k_I\sim{\cal O}(1)$.

Finally, in Case (iii), we have
\be\label{eqn:case iii criterion}
A\sim\g^{w-1}
\K^{1+w}\k^w\hbar^{1+w}\D^\frac{w-1}{2},
\ee
which yields
\ba
p_{I,{\rm crit}}\sim p_{\rm crit}.\\\nn
\ea
Therefore, in Case (iii), the big bang singularity is replaced by big bounces as well and the bouncing points of $p_I$ are between $p_{I,{\rm crit}}$ given by \eqnref{eqn:p_I crit} and $p_{\rm crit}$ given by \eqnref{eqn:p crit} ($p_{I,{\rm crit}}$ and $p_{\rm crit}$ are now in the same order).

To sum up, the big bang singularity is replace by big bounces due to the fact that the gravity with loop quantum corrections becomes repulsive at some point when the universe is near
the singularity. In Cases (i), this happens when the directional density $\vr_I$ reaches Planckian energy density $.86\r_{\rm Pl}$; in Case (ii), it happens when the matter density $\r_M$ approaches $\approx.86\r_{\rm Pl}$. In all cases, the big bounce takes place whenever either of $\vr_I$ or $\r_M$ approaches ${\cal O}(\r_{\rm Pl})$ first. In a sense, there is competition between the energy density carried from the classical anisotropic shear ($\vr_I$) and the matter density ($\r_M$) to be the indication of occurrence of the big bounces.

Also notice that as mentioned in \secref{effective theory}, $\Sigma^2(\text{post bounce})\neq\Sigma^2(\text{pre bounce})$ in general due to the fact that the anisotropies are smeared by the quantum corrections. However, in Case (i), the information of anisotropies persists during the bouncing period and we shall have $\Sigma^2(\text{post bounce})\approx\Sigma^2(\text{pre bounce})$. This can be seen from \eqnref{eqn:bouncing p_I}, in which the evolutions for different $p_I$ are decoupled around the bounces and therefore the constants $\K_I$ (which dictate anisotropies) are unchanged before and after the bounce.

The equations of motion given by \eqnref{eqn:qm dp/dt'} and \eqnref{eqn:qm dc/dt'} can be solved numerically once the initial values $p_I(t=t_0)$ and $c_I(t=t_0)$ are given. Equivalently, for given $p_I(t_0)$, specifying $\K_I$ is an alternative way to specify $c_I(t_0)$. [Given $\K_I$, $f(t_0)$ can be obtained via \eqnref{eqn:f(t')} and then $c_I(t_0)$ are fixed by \eqnref{eqn:K+f} if the initial condition is in the classical regime and by \eqnref{eqn:qm G_I} if in quantum regime.] By changing the parameters $A$ and $\K$, we can get the three different cases as discussed above. The numerical solutions in the presence of radiation are solved in terms of proper time $t$ and depicted in Figures \ref{fig:case1}, \ref{fig:case2} and \ref{fig:case3} for these three cases respectively.

\begin{widetext}

%=============================================================================
%=======  Figure for Case (i):
%-----------------------------------------------------------------------------
\begin{figure}
\begin{picture}(400,140)(0,0)
%\graphpaper(0,0)(400,150)

\put(-60,-10){
\begin{picture}(460,140)(0,0)
%\graphpaper(0,0)(460,150)

\resizebox{\textwidth}{!}
{\includegraphics{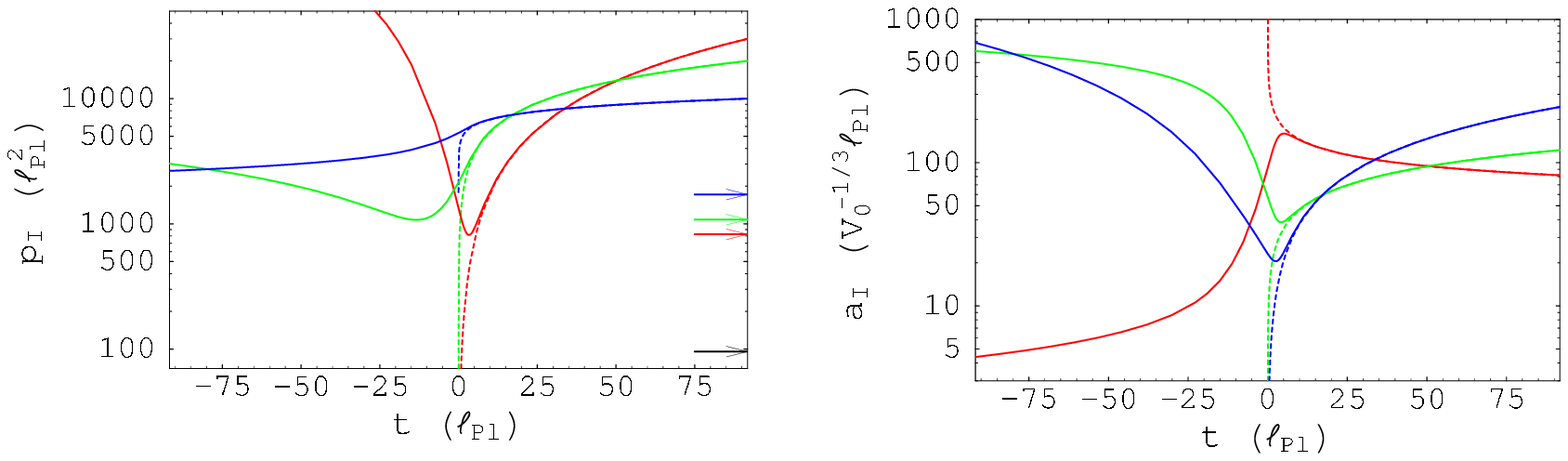}}

\end{picture}
}

\end{picture}
\caption{\textbf{Case (i): Quantum corrections take effect in the Kasner phase.} With $w=1/3$ (radiation filled);
$\k_1=-2/7$, $\k_2=3/7$, $\k_3=6/7$;
$\K=1.\times10^3$; $A=1.\times10^2\hbar\,\Pl^{3w-1}$; and $p_1(t_0)=3.\times10^4\Pl^2$,
$p_2(t_0)=2.\times10^4\Pl^2$,
$p_3(t_0)=1.\times10^4\Pl^2$.
(Also set $\g=1$.)
The
%\textcolor{red}
{red} curves are for $p_1$, $a_1$;
%\textcolor{green}
{green} for $p_2$, $a_2$;
and
%\textcolor{blue}
{blue} for $p_3$, $a_3$. Solid lines are the solution to the effective loop quantum evolution; dashed lines to the classical evolution. The proper time $t$ is offset such that the classical singularity is at the origin of $t$. The values of $p_{I,{\rm crit}}$ are pointed by the colored arrows and $p_{\rm crit}$ by the black one. In this case, the bouncing point of each $p_I$ matches $p_{I,{\rm crit}}$ very precisely and we have $p_{I,{\rm crit}}\gg p_{\rm crit}$ (or say $\r_M\ll\vr_I\sim\r_{\rm Pl}$ near the bounces). [Note that the bounce of $p_3$ is out of the shown range. The isotropized phases on both sides of the bounces are also out of plot.]}\label{fig:case1}
\end{figure}

%=============================================================================
%=======  Figure for Case (ii):
%-----------------------------------------------------------------------------
\begin{figure}
\begin{picture}(400,140)(0,0)
%\graphpaper(0,0)(400,150)

\put(-60,-10){
\begin{picture}(460,140)(0,0)
%\graphpaper(0,0)(460,150)

\resizebox{\textwidth}{!}
{\includegraphics{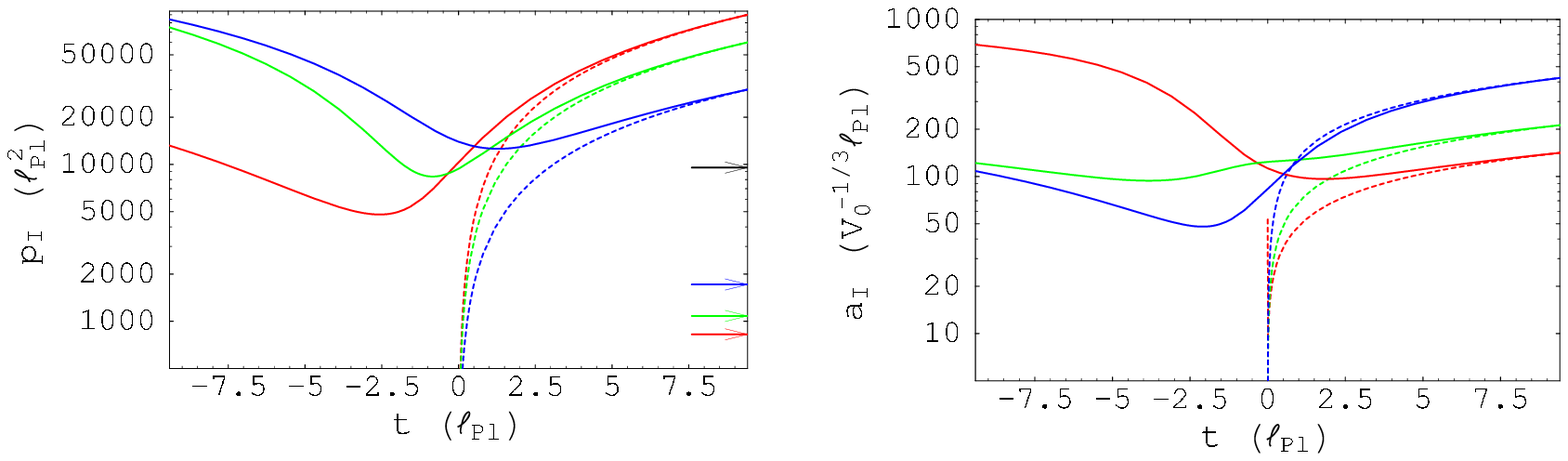}}

\end{picture}
}

\end{picture}
\caption{\textbf{Case (ii): Quantum corrections take effect in the isotropized phase.} With $w=1/3$ (radiation filled);
$\k_1=-2/7$, $\k_2=3/7$, $\k_3=6/7$;
$\K=1.\times10^3$; $A=1.\times10^6\hbar\,\Pl^{3w-1}$; and $p_1(t_0)=9.\times10^4\Pl^2$,
$p_2(t_0)=6.\times10^4\Pl^2$,
$p_3(t_0)=3.\times10^4\Pl^2$.
In this case, the bouncing point of each $p_I$ roughly matches $p_{\rm crit}$ and we have $p_{I,{\rm crit}}\ll p_{\rm crit}$ (or say $\vr_I\ll\r_M\sim\r_{\rm Pl}$ near the bounces). [Note that in the backward evolution the contracting curve of the classical $a_1$ eventually becomes expanding at the epoch extremely close to the singularity, indicating that the Kasner phase does occur classically although almost invisible in the plot.]}\label{fig:case2}
\end{figure}

%=============================================================================
%=======  Figure for Case (iii):
%-----------------------------------------------------------------------------
\begin{figure}
\begin{picture}(400,140)(0,0)
%\graphpaper(0,0)(400,150)

\put(-60,-10){
\begin{picture}(460,140)(0,0)
%\graphpaper(0,0)(460,150)

\resizebox{\textwidth}{!}
{\includegraphics{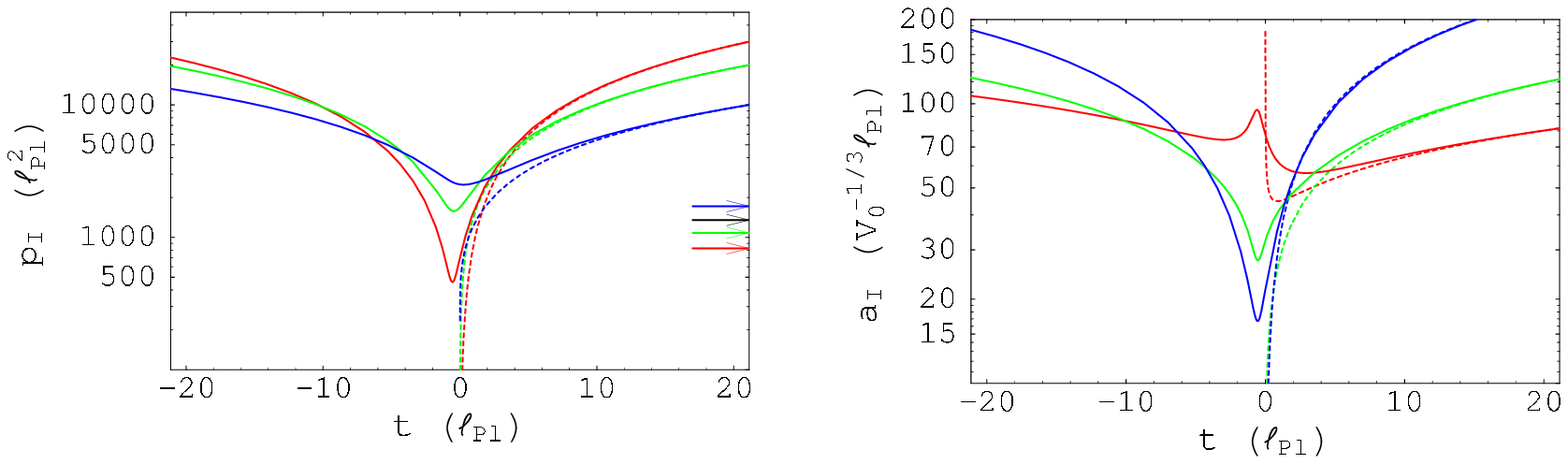}}

\end{picture}
}

\end{picture}
\caption{\textbf{Case (iii): Quantum corrections take effect in the transition phase.} With $w=1/3$ (radiation filled);
$\k_1=-2/7$, $\k_2=3/7$, $\k_3=6/7$;
$\K=1.\times10^3$; $A=2.\times10^4\hbar\,\Pl^{3w-1}$; and $p_1(t_0)=3.\times10^4\Pl^2$,
$p_2(t_0)=2.\times10^4\Pl^2$,
$p_3(t_0)=1.\times10^4\Pl^2$.
In this case, we have $p_{I,{\rm crit}}\sim p_{\rm crit}$, roughly around which all $p_I$ are bounced (or say $\vr_I\sim\r_M\sim\r_{\rm Pl}$ near the bounces).}\label{fig:case3}
\end{figure}

\end{widetext}

\section{Alternate Quantization} \label{altmubar}
In this section we consider the effective dynamics of an alternative quantization
scheme mentioned in \secref{effective theory}. The major difference is that the behavior
of the parameters $\mubar_I$ varies with the triad components differently. Specifically
we take the dependence as $\mubar_I \propto 1/a_I$ with the exact behavior for
$\mubar_1$ given as
\be
	\mubar_1 = \sqrt{\frac{\D p_1}{p_2 p_3}}\ .
\ee

The effective Hamiltonian is similar in form to that in \eqref{eqn:qm Hamiltonian}
and is given explicitly by
\ba\label{eqn:qm Hamiltonian2}
\Ham_{\rm eff}&=&
-\frac{1 }{\k \g^2}
\bigg\{
 \frac{\sin(\mubar_2c_2)\sin(\mubar_3c_3)}{\mubar_2 \mubar_3}\,p_2p_3\nn\\
&&\quad+\text{cyclic terms}
\bigg\}
+\,p_1p_2p_3\,\rho_M,
\ea
which becomes
\ba
\Ham_{\rm eff}&=&
-\frac{p_1p_2p_3 }{\k \g^2 \D}
\bigg\{
\sin(\mubar_2c_2)\sin(\mubar_3c_3)\nn\\
&&\quad+\text{cyclic terms}
\bigg\}
+\,p_1p_2p_3\,\rho_M.
\ea

With this, the vanishing of the constraint leads to the simple relation
\be
	\rho_M = \frac{1}{\k \g^2 \D}
	\bigg\{
	\sin(\mubar_2c_2)\sin(\mubar_3c_3)\nn\\
	+\text{cyclic terms}
	\bigg\}
\ee
and from this we can deduce an important observation. Namely, since
all of the sin terms are bounded we find that the energy density is also
bounded as
\be
	\rho_M < \frac{3}{\k \g^2 \D} \equiv \rho_{\rm crit},
\ee
where $\rho_{\rm crit}$ is numerically the same value as in the
isotropic case
\be
	\rho_{\rm crit} \equiv \frac{3}{\k \g^2 \D} \approx .82 \rho_{\rm Pl}.
\ee
The fact that $\rho_M$ is bounded implies that the big-bang singularity is resolved since
classically the energy density blows up there. Note that any bounce does not necessarily occur
when $\rho_M = \rho_{\rm crit}$ and we shall see, that with anisotropies the bounce occurs at lower energy densities.

We can derive Hamilton's equations in the same fashion as in \secref{effective theory}
and it can be shown that
\be\label{pIcI}
	p_I c_I - p_J c_J =  \g \Vz \, \a_{IJ},
\ee
where $\alpha_{IJ}$ are constants of motion. Note that this exact relation
was satisfied in the classical case in \eqref{aIJ}. However, with this
effective Hamiltonian, $c_I$ are not as simply related to $\dot{a}_I$ as
in the classical case given in \eqref{eqn:cl a dot}. The full relation
is to be determined from the Hamilton's equation for $\dot{p_I}$ which
give for instance
\be \label{pdot}
	\dot{p_1} = \frac{p_1}{\sqrt{\D} \g} \cos(\mubar_1c_1)\bigg(\sin(\mubar_2c_2)+\sin(\mubar_3c_3)\bigg).
\ee
Because of this, the shear term $\Sigma^2 = \frac{a^6}{18} [ (H_1-H_2)^2 + (H_2-H_3)^2+(H_3-H_1)^2]$ is no longer constant. However, in the classical limit where $\mubar_I c_I \ll1$,
we can calculate the shear term to be $\Sigma^2 \approx \frac{1}{18} \big( \alpha_{12}^2 + \alpha_{23}^2
	+ \alpha_{31}^2 \big)$ as in the classical case.

Therefore if we consider the behavior of the shear term and begin with a semi-classical
collapsing universe, we find that initially $\Sigma^2$ is constant with value
\be
	\Sigma^2 \approx \bar{\Sigma}^2 \equiv \frac{1}{18} \big( \alpha_{12}^2 + \alpha_{23}^2
	+ \alpha_{31}^2 \big).
\ee
Through the bounce $\Sigma$ is not constant, but after the bounce in the
expanding phase where once again $\mubar_I c_I \ll1$, we find
that the shear takes on the same value. This is because equation
\eqref{pIcI} holds throughout the evolution. Therefore it is a rather general feature
of this effective Hamiltonian, that the shear term $\Sigma$ is conserved before and
after the collision. Note that this is in contrast with the results presented in
the body of this paper, where $\Sigma$ was only conserved for a massless scalar
field. However, both quantizations imply that the shear term is bounded and
the anisotropies do not blow up.

Because of the complexity of the equations of motion, it is highly non-trivial
to derive a generalized Freidmann equation. We can however if we expand the
effective Hamiltonian to second order in $\mubar_I c_I$. Then  equations
\eqref{pIcI} and \eqref{pdot} can be used to derive an approximate
generalized Friedmann equation given by
\ba
	H^2 &=& \frac{\kappa}{3} \rho_M \Big( 1 - \frac{\rho_M}{\rho_{\rm crit}}\Big)
	+ \frac{\bar{\Sigma}^2}{a^6} 	\nn \\
	&&- \frac{3 \bar{\Sigma}^2 \rho_M}{a^6 \rho_{\rm crit}} - \frac{9 \bar{\Sigma}^4 }{\kappa a^{12} \rho_{\rm crit}} + \mathcal{O} \big((\mubar_I c_I)^4\big).
\ea
If we solve this for the matter energy density at the bounce ($H=0$) we find
\ba
\rho_{\rm bounce}
&\approx&1/2\Bigg( \rho_{\rm crit} - \frac{9 \bar{\Sigma}^2 }{\kappa a^6 } \nn \\ &&+ \sqrt{(\rho_{\rm crit}-\frac{9 \bar{\Sigma}^2 }{\kappa a^6 }) (\rho_{\rm crit}-\frac{3 \bar{\Sigma}^2 }{\kappa a^6 })}\;\Bigg),\qquad
\ea
which implies that the energy density is bounded below $\rho_{\rm crit}$ in accordance
with the prediction from the vanishing of the effective Hamiltonian. Note that
$\bar{\Sigma}^2$ is a constant of motion and only related to the shear
term $\Sigma^2$ in the classical limit $\mubar_I c_I \ll 1$. Note also, that this
generalized Friedmann equation is valid in the nearly isotropic limit which can be understood from the relations \eqref{pIcI}
which imply
\be
	\mubar_I c_I - \mubar_J c_J = \frac{\g \sqrt{\D} \alpha_{IJ}}{a^3}.
\ee
If the right hand side is not small (i.e., when the anisotopies are large), then
at least one $\mubar_I c_I$ is guaranteed to be large and hence
the generalized Friedmann equation would need to be calculated to higher order
to provide a good approximation.

We note that both the quantization scheme presented in this appendix and that given
in the body of this work agree as to the qualitative nature of the bouncing
universe. Namely, a bounce still occurs with the inclusion of anisotropies and that the shear term does not blow up implying that  the anisotropies remain finite through the bounce.

%\bibliography{../lqc}
%\bibliographystyle{unsrt}

\end{document}